# The curious case of A31P, a topology-switching mutant of the Repressor of Primer protein : A molecular dynamics study of its folding and misfolding

Olympia-Dialekti Vouzina, Alexandros Tafanidis

& Nicholas M. Glykos*

Department of Molecular Biology and Genetics, Democritus University
of Thrace, University campus, 68100 Alexandroupolis, Greece, Tel +30-25510-30620,
Fax +30-25510-30620, https://utopia.duth.gr/glykos/ , glykos@mbg.duth.gr

# Abstract

The effect of mutations on protein structures is usually rather localized and minor. Finding a mutation that can single-handedly change the fold and/or topology of a protein structure is a rare exception. The A31P mutant of the homodimeric Repressor of Primer (Rop) protein is one such exception: This single mutation —and as demonstrated by two independent crystal structure determinations— can convert the canonical (left-handed/all-antiparallel) 4-α-helical bundle of Rop, to a new form (right-handed/mixed parallel and antiparallel bundle) displaying a previously unobserved 'bisecting U' topology. The main problem with understanding the dramatic effect of this mutation on the folding of Rop is to understand its very existence : Most computational methods appear to agree that the mutation should have had no appreciable effect, with the majority of energy minimization methods and protein structure prediction protocols indicating that this mutation is fully consistent with the native Rop structure, requiring only a local and minor change at the mutation site. Here we use two long (10 μs each) molecular dynamics simulations to compare the stability and dynamics of the native Rop *versus* a hypothetical structure that is identical with the native Rop but is carrying this single Alanine$_{31}$ to Proline mutation. Comparative analysis of the two trajectories convincingly shows that in contrast to the indications from energy minimization —but in agreement with the experimental data—, this hypothetical native-like A31P structure is unstable, with its turn regions almost completely unfolding, even under the relatively mild 320K *NpT* simulations that we have used for this study. We discuss the implication of these findings for the folding of the A31P mutant, especially with respect to the proposed model of a double-funneled energy landscape.

# 1 Introduction

The Repressor of primer (Rop) protein is the paradigm of a canonical homodimeric 4-α-helical bundle (see Fig.1, upper panel). Ever since its genetic identification by Twigg & Sherratt more than 40 years ago,[1] it has been studied exhaustively in terms of its genetics,[2-4] molecular biology,[5-8] biochemistry,[9-16] structure,[17-34] folding,[35-49] and, more recently, of its complex sequence/structure/folding relationships and of its applications in protein design.[50-70] The combination of all those studies, makes Rop one of the best characterized 4-α-helical bundles known today.

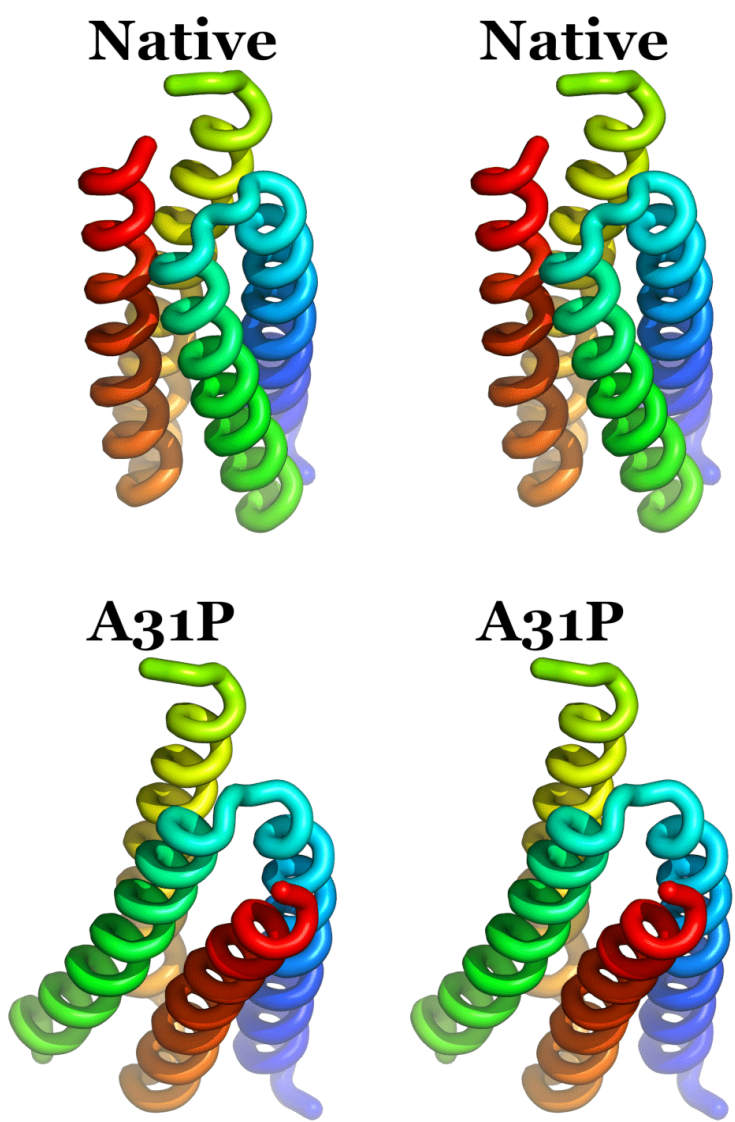

**Fig.1** Wall-eyed stereodiagrams of the crystallographic structures of native Rop (upper panel) and A31P (lower panel). The coloring scheme for the two structures is the same and ranges from blue for the N-terminus of first monomer, to red for the C-terminus of the second monomer. The two structures have been superimposed on the N-terminal helix of the first monomer (colored blue/cyan).

Numerous mutants and variants of Rop have been studied over the last three decades, mainly with the aim of elucidating its complex sequence/structure relationships.[50–70] In the course of those studies, Rop demonstrated a remarkable structural plasticity with several different topologies and oligomerization states being experimentally characterized for different mutants and variants (see, for example, Fig.1 of Glykos *et al*[29]). A persistent issue with the majority of those studies was the absence of a folding perspective: connecting a given Rop variant sequence with its corresponding structure is satisfactory, but it lacks the deeper understanding that only a folding study can provide. The reason for this absence is that studying the folding of Rop is difficult. The combination of a symmetric homodimeric protein, together with its very slow folding dynamics,[46,47] and the (assumed) presence of frustrated and complex folding landscapes,[44,45] made folding studies of Rop rather rare and mostly qualitative.

The most intensively studied part of the Rop structure is the turn region connecting the two α-helices of each monomer. Numerous mutants and variants of the turn residues have been designed, expressed, and characterized thermodynamically, kinetically and structurally,[12,13,25,29,38–41,43,49,50,52,54,57,65] mainly with the aim of understanding the role of turns in the stability, structure and folding of the bundle. One of the most structurally impressive —and least understood— mutants of Rop is the A31P mutant.[13,24,25,27,43,44] Fig.1 shows a comparison of the native Rop structure (upper panel) with the crystallographic structure of the A31P mutant (lower panel). The mutant structure is completely reorganized and is converted from the canonical left-handed/all-antiparallel 4-α-helical bundle of native Rop, to a right-handed, mixed parallel and antiparallel bundle, displaying a 'bisecting U' topology. This change of topology is accompanied by a complete re-organization of the structure at the atomic level, including the mutant's hydrophobic core. Whereas the native Rop hydrophobic core shows the typical —for coiled-coils— motif of successive layers of interacting residues of the type *adad* where *a* and *d* are the apolar positions of the heptad repeat characterizing coiled-coils, A31P demonstrates an highly heterogeneous collection of interactions[25] which includes residues of the type *dddd*, *ggaaa* and *gdd*. The result of these changes is a significant destabilization of the A31P structure compared with the native Rop.[13]

The most fundamental problem with understanding the structural effects of the A31P

mutant, is that according to current structure prediction and modeling methods, this mutation should not have had any appreciable effect on the structure of Rop. As will be discussed extensively in the next section, a native Rop-like structure of A31P looks entirely normal to both computational methods and trained human observers alike.

In this communication we attempt to answer the following question : Why is A31P not native-like ? Is this mutation really inconsistent with the native Rop structure ? We tackle this question by comparing an extensive molecular dynamics simulation of native Rop *versus* a simulation of a hypothetical native-like structure of A31P as produced by structure prediction software. We present evidence that the A31P mutation —and in contrast to the initial modeling indications— appears to be incompatible with the native Rop structure, with its turns initiating cycles of unfolding and refolding even under the mild simulation conditions used for this study. We close by discussing the implications of these findings for the folding of Rop and its A31P mutant.

## 2 Modeling and structure prediction software both suggest a native-like structure for A31P

The first indication that the A31P mutant appeared to be compatible with the native Rop structure —requiring only a minor relaxation at the mutation site— came from modeling attempts that were performed long before the actual crystal structure determination was reported.[25] Fig.2 shows an example of how Proline$_{31}$ may be fitted in the loop of the native Rop structure without causing steric clashes or other obvious problems. This initial indication was corroborated by the results obtained from standard energy minimization methods of protein structures. Using, for example, the GalaxyWEB server[71,72] to perform energy minimization of native-like models of A31P (prepared with VMD,[73] Pymol[74] and Coot[75]), gave refinement energies that were quite similar to the energies obtained from the native structure (-6085±13 for native Rop *versus* -6013±12 for the native-like A31P (these averages and standard deviations were calculated from the best five models that GalaxyWEB prepares by default in each run)). When the refinement energy of the real (bisecting U) structure of A31P was calculated, it was found to be significantly higher at -5632±11, in good agreement with the experimentally known destabilization of A31P.[13] These results immediately demonstrate the *paradox* that the structure of this mutant poses: if a native Rop-like structure of A31P is energetically more favorable than its real (crystallographically determined) structure, then A31P should not have folded as it does. We see two ways out of this paradox. The first is that A31P never visits a native-like structure during its folding. Given that the folding of Rop is extremely slow (of the order of seconds[46,47]), it appears highly unlikely that a native-like structure is never sampled during A31P folding. The second solution is that the refinement energies do not tell us the whole story (because, for example, they miss entropic contributions), and that in reality a native-like structure for A31P is unstable.

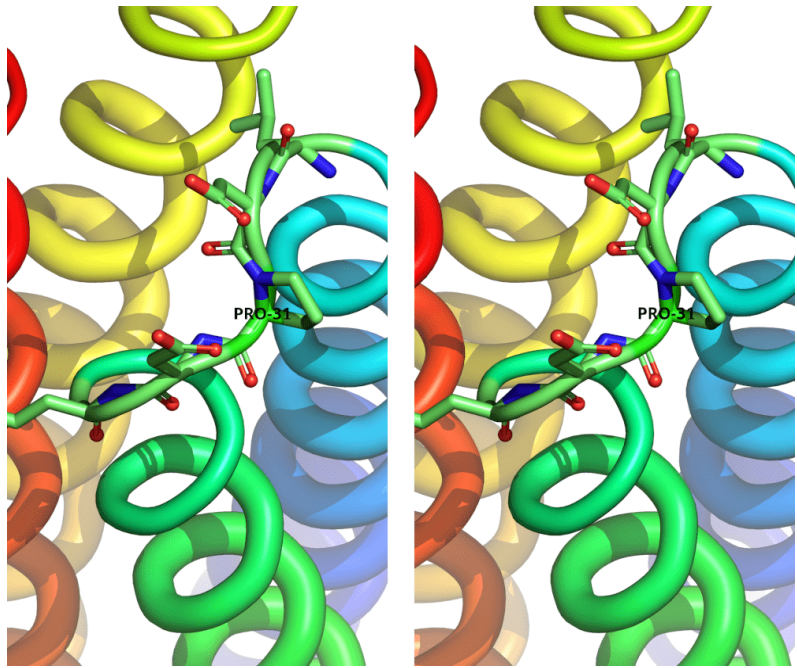

**Fig.2** Wall-eyed stereodiagram of a hypothetical native-like A31P structure. The coloring scheme for the two monomers is the same with the one shown in Fig.1. Residues 29-33 are depicted in liquorice representation, and $Pro_{31}$ is labeled.

To further examine the validity of the proposition that a native-like structure for the mutant appears to be entirely plausible based on modeling methods, we have used AlphaFold[76–78] to prepare structure prediction models for native Rop and five of its turn mutants and variants, including A31P. This calculation was performed using the defaults of the AlphaFold2-multimer ColabFold v1.5.5 interface available via

colab.research.google.com. The aim of this modeling exercise was three-fold. The first was to apply a fully automated and verifiable procedure, removing all doubts which may have arisen from using interactive modeling methods based on molecular graphics programs such as Pymol and Coot. The second was mostly curiosity driven: would AlphaFold select as a model structure the crystallographically known A31P structure (with 100% sequence identity), or the weight of all those native-like sequences deposited with the PDB would revert the algorithm to a Rop-like structure ? Last, but not least, by selecting mutants which also included a proline mutation at the turn region (but which were known experimentally to be native-like), we could directly compare the weight that the prediction software would give to a proline at position 31 *versus* —for example— a proline at position 30.

The variants and mutants for which AlphaFold structure prediction models were prepared (using as sole input their sequences and their dimeric oligomerization state) are:

1. Native Rop.
2. The D30P mutant which is known from the experiment[54] to have a native-like structure.
3. The E28A-D30P-D32A mutant, hereafter referred to as the "APA" mutant. This mutant is also known from the experiment[54] to have a native-like structure.
4. The D30P-A31G mutant, hereafter referred to as the "PG" mutant. This mutant is also known from the experiment[68] to have a native-like structure.
5. The A31P mutant.
6. The "2AA" variant. This is a double insertion variant that restores Rop's heptad repeat at the turn region. The insertion comprises two alanine residues inserted immediately before and immediately after $Asp_{30}$ (in the native Rop numbering). This variant is also known experimentally[50] to be native-like.

Please do note that with the exception of A31P, all other sequences in this list are known experimentally to have a native Rop-like structure, with three of them displaying a proline mutation but at position 30 (instead of 31). We perceive this as an additional reason that makes the consequences of the A31P mutation so difficult to grasp.

Table I shows the RMSD values (calculated with `MMalign`[79]) between all possible pairs of AlphaFold-derived structures for the six sequences described above. For reference, the crystallographic structure of native Rop (marked as "Exp") is also included in the RMSD matrix.

**Table I** RMSD values between all pairs of AlphaFold-modelled mutants, see text for details. The crystallographically determined Rop structure (marked as "Exp") and the AlphaFold-derived model of the native Rop (marked as "Nat"), are also included. All values are in Ångström, and only the Cα atoms were used for this calculation.

|      | Exp | Nat  | D30P | APA  | PG   | A31P | 2AA  |
|------|-----|------|------|------|------|------|------|
| Exp  | —   | 0.49 | 0.68 | 0.77 | 0.84 | 0.67 | 0.73 |
| Nat  |     | —    | 0.47 | 0.57 | 0.69 | 0.49 | 0.60 |
| D30P |     |      | —    | 0.21 | 0.44 | 0.47 | 0.38 |
| APA  |     |      |      | —    | 0.41 | 0.52 | 0.43 |
| PG   |     |      |      |      | —    | 0.52 | 0.48 |
| A31P |     |      |      |      |      | —    | 0.52 |
| 2AA  |     |      |      |      |      |      | —    |

The main message of this modeling exercise with AlphaFold is immediately obvious even at this stage of the analysis: All Rop variants and mutants studied here are predicted to have a native Rop structure, including the A31P mutant. The fact that AlphaFold predicts a native-like structure for A31P although there are two PDB entries for this mutant displaying the 'bisecting U' topology (entries 1B6Q and 1GMG), could be considered a failure of the algorithm. This is not the case: AlphaFold is fundamentally an evolution-based approach and is not suitable for predicting structures for sequences with no evolutionary history (like those Rop mutants and variants). It is only natural that the weight of tens of Rop-like sequence/structure pairs was given precedence over the two A31P entries (especially considering that native Rop and A31P share ~98% sequence identity).

The RMSD matrix of Table I is useful for comparing the final predicted structures, but carries no information about the estimated errors of the modeling procedure. These estimated errors (as produced by the AlphaFold interface) are shown in Fig.3.

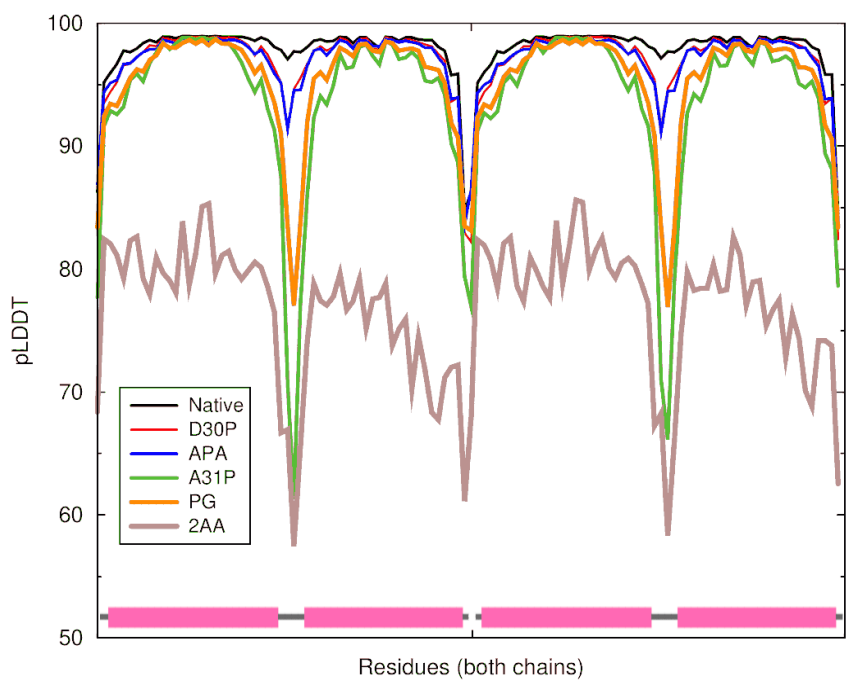

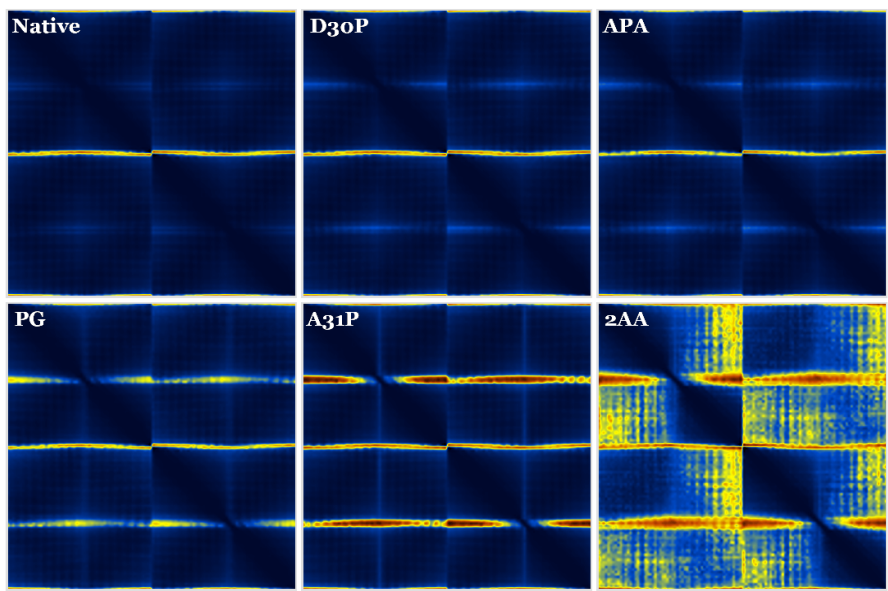

**Fig.3** Statistics for the AlphaFold models. Panel (A) depicts the variation of the predicted local-distance difference test (pLDDT) as a function of residue number for all mutants and variants studied in this report. The schematic at the lower part of the diagram indicates the positions of the helices (depicted as cylinders) and turns of the two monomers. Panel (B) shows the predicted aligned error (PAE) maps, which are a measure of the quality of multimeric modeling. In each of these maps the origin is at the upper left-hand corner, the vertical and horizontal axes correspond to successive residues of both chains, and the predicted aligned error values range from zero (dark blue) to 27Å (dark red). The four smaller dark squares along the diagonal of each map (most easily seen in the 2AA diagram) correspond to the four helices of the structures.

Fig.3A shows the variation of the predicted local-distance difference test (pLDDT) as a function of residue number for all six sequences studied here. pLDDT is a measure of confidence in the per-residue modeling of the target sequence, with its normalized values range from 0 to 100. pLDDT carries no information about the confidence in the relative placement of the individual helices and monomers,[80] but it clearly does show that AlphaFold correctly identifies the turns as the areas where the least confidence must be placed. Having said, with the exception of the 2AA variant, all AlphaFold-produced models have average pLDDT scores that are in the 'safe modeling' zone as shown in Table II.

**Table II** Average pLDDT, pTM, and ipTM values for the best refined models obtained from AlphaFold for all Rop variants and mutants studied in this report. pLDDT scores range from zero to 100 and are a measure of the local structural accuracy. pTM/ipTM scores are in the range 0.0 to 1.0 and consistute a measure of topological accuracy for the case of multimeric modeling, see text for details.

|       | Nat   | D30P  | APA   | PG    | A31P  | 2AA   |
|-------|-------|-------|-------|-------|-------|-------|
| pLDDT | 97.8  | 97.0  | 96.9  | 95.2  | 93.7  | 77.1  |
| pTM   | 0.909 | 0.902 | 0.896 | 0.882 | 0.885 | 0.677 |
| ipTM  | 0.898 | 0.887 | 0.882 | 0.867 | 0.867 | 0.641 |

The same overall picture emerges when examining quantitative measures of the estimated topological accuracy of the dimeric models. The average pTM/ipTM scores[80] (Table II) are very similar for all mutants and variants, and almost identical with the scores obtained from native Rop, indicating a confident topological modeling. These findings are in agreement with the predicted aligned error maps shown in Fig.3B: the relative placement of the helices and monomers is modeled confidently for all mutants and variants with the exception of the double-insertion 2AA variant. Higher values of the PAE score are only observed for the turns and the terminal residues, with all other inter-helical cross terms being in the ~1-2Å range.

To summarize this section, both modeling and structure prediction/refinement methods indicate that a native-like structure for A31P is not just feasible, but that it appears to be as good a structure as for several other Rop mutants and variants that are known experimentally to be native-like.

# 3 Molecular dynamics simulations

## 3.1 Aim and limitations

We have performed extensive (10 μs each) molecular dynamics simulations in the *NpT* ensemble (T=320K) of (a) the native Rop structure, and, (b) of a native-like A31P structure as discussed in the previous section and shown in Fig.2. The aim of these simulations was to allow us to establish (through their comparison) whether a native Rop-like structure for A31P is indeed as stable as the modeling and structure prediction methods indicated (see previous section). Before continuing with the presentation and analysis of these simulations we must first discuss the ever present issue of convergence and sufficient sampling of the corresponding trajectories. Rop and most of its variants and mutants are very slow folders, and are known to be even slower to unfold.[46,47] With folding times of the order of tens of seconds, and even longer unfolding times, the question of sufficient sampling of folding simulations is outside present day computing capabilities. The important point, however, is that our simulations are not folding simulations, but are initiated with the proteins in the folded state. The implication is that if the structures of both the native and the A31P trajectories are stable, then we could indeed observe sufficiently sampled dynamics (of their folded state), and we could confidently reject the hypothesis that a native-like A31P structure is not stable. If on the other hand, the A31P structure is not stable and it initiates even partial unfolding, then sufficient sampling of its unfolding is clearly out of the question, *but* in the context of this study this deficiency is irrelevant: Our aim is not to study the unfolding of this hypothetical structure, but to clarify whether it is as stable as the modeling and structure prediction calculations indicated. Quantifying sufficient sampling of these two trajectories is discussed later in this section.

## 3.2 Simulation protocol

The simulation protocol used in this study is essentially identical with the one previously reported by this group (see, for example, Gkogka & Glykos[81]), and will not be discussed in detail. In summary, it is a classical *NpT* simulation performed at 320K and 1 atm using the AMBER99SB-STAR-ILDN force field,[82,83] periodic boundary conditions, explicit representation of the solvent using the TIP3P water model,[84] full PME-based electrostatics, and a 2.5fs timestep. The systems were prepared with the `leap` module of AMBER tools,[85] the simulations performed with the program NAMD,[86] and the trajectories analyzed with the programs `carma`[87] and `grcarma`.[88]

Although this is a standard and more-or-less fully established protocol, there are two aspects of its implementation that must be discussed. The first concerns the usual suspect, the choice of force field. We have elected to use the AMBER99SB-STAR-ILDN force field for two reasons. The first, and as demonstrated by Shaw and coworkers,[89] is that this force field showed the closest agreement with the experimental chemical shifts and NOEs in the dataset by Mao *et al.*[90] comprising 41 folded proteins, and the most stable simulations in the data set of Huang *et al.*,[91] comprising 11 folded proteins. The second reason is that this force field not only is one of the best for the study of proteins in the folded state, but to our knowledge and experience is also one of the best performing force fields for folding simulations of peptides and small proteins.[81,92–99] The known issue with the too compact unfolded states produced by this force field[100] is mostly irrelevant for this study —which is initiated with proteins in the folded state.

The second aspect of the simulations that must be discussed is the absence of an enhanced sampling method,[101] like, for example, adaptive tempering[102] (which we have used in most of the folding studies we have previously reported[81,92–99]). The reason for our choice not to use such a method is more subtle: Most enhanced sampling methods are equivalent to modifying, directly or indirectly, the energy landscape of the respective systems. We believe that inclusion of an enhanced sampling method would invalidate our attempt to be able to directly and immediately compare the two trajectories without the complications and putative sources of additional systematic errors introduced by the enhanced sampling method *per se*. To give a more solid example of the problem, consider the adaptive tempering method which is equivalent to a single-copy replica exchange simulation with a continuous temperature range (applied through the Langevin thermostat). If we had implemented this method, then the two simulations would necessarily have sampled different temperature distributions during the finite time of the simulations. This would invalidate our attempt to directly compare the two trajectories because there would be no obvious way to accurately refer the two simulations to a reference temperature distribution. It is for this reason that we elected to perform two relatively long, but "safe", *NpT* simulations at a temperature where the native Rop fold is expected to be structurally stable.

## 3.3 Sufficient sampling: Good-Turing estimates

We have applied Good-Turing statistics[103] to estimate the probability of observing significantly different structures of the two molecules should the simulations were extended to longer time intervals. The results, using only the turn residues of the respective structures, are shown in Fig.4 in the standard (for this method) form of *(P vs. RMSD)* graphs.

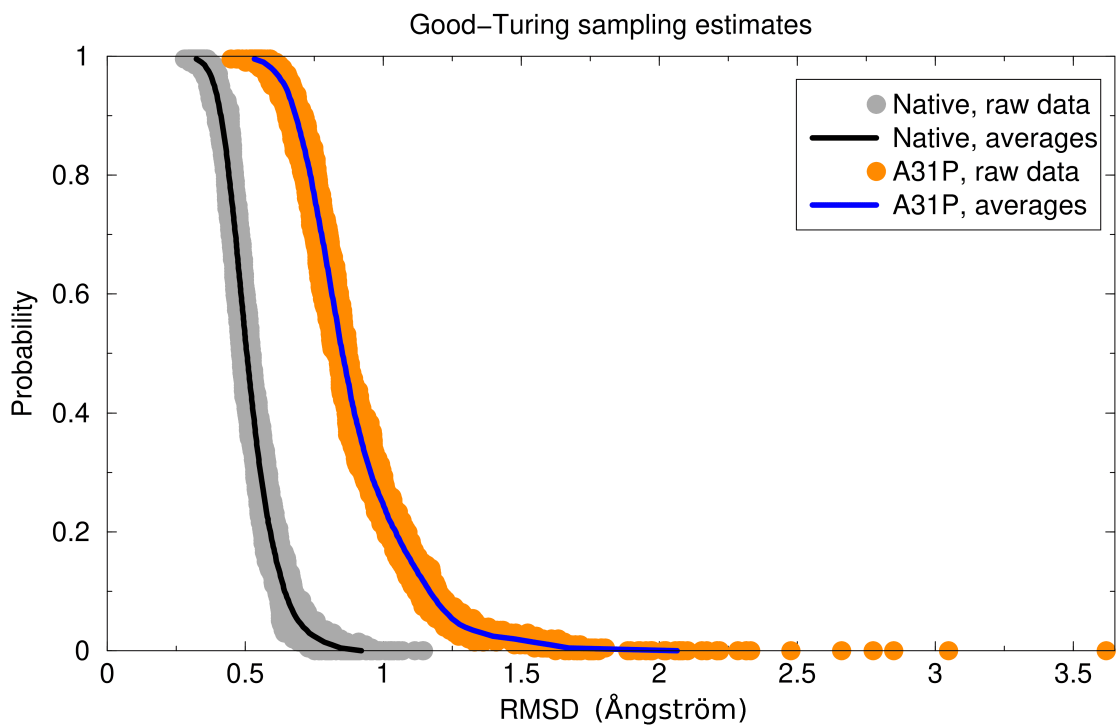

**Fig.4** Good-Turing sampling estimates for the two trajectories using the turn residues 24-39 of the two structures. Both the raw data and the averages are depicted with the color coding indicated in the insert. Notice the A31P raw data points reaching out to an RMSD of ~3.6Å. See text for a discussion of these graphs.

We will give a numerical example to clarify the information content of these graphs. Looking at the A31P curve, an RMSD value of 1.0Å corresponds to a probability value of approximately 0.25. This means that if we were to continue the simulation, we should expect that ~25% of the new (previously unobserved structures) would differ by an RMSD of at *least* 1.0Å from all structures that we already observed (in the 10 μs trajectory). If we now examine the native Rop graph, for an RMSD of 1.0Å the corresponding probability is less than 0.005, which implies that if we were to extend the native simulation, less than 0.5% of the new structures would have an RMSD of 1.0Å or more from the already observed native structures.

We can reach the same conclusions about the stability and sampling of the two trajectories by examining the maximal RMSDs of these graphs as shown by the raw data (orange and gray filled circles in Fig.4). The native Rop simulation shows a maximal RMSD of ~1.2Å (rightmost gray circle) versus an RMSD of ~3.6Å for A31P. Given that the quoted RMSDs were calculated using the turn residues only, these results clearly indicate the presence of significant differences in the stability of the two structures.

In summary, the Good-Turing analysis indicates that the native Rop structure appears to be exceedingly stable during the simulation, and its dynamics are sufficiently sampled. In contrast, the turn of the A31P structure appears to be unstable and as will be shown conclusively in the next section, this is due to its partial unfolding.

## 3.4 Direct structural comparison illustrates the extent of unfolding of the A31P turn

Fig.5 shows a direct comparison between the native and A31P structures which showed the highest RMSD from their respective starting conformations (1.76 and 3.62Å respectively for residues 24-39, see Fig.6 for the actual *RMSD vs. time* diagrams). Taken together Figures 5 and 6 convincingly indicate that the A31P mutation is incompatible with the native Rop fold, mainly because it destabilizes and subsequently initiates the unfolding of the turn regions, leading to the exposure and subsequent dissolution of the hydrophobic core. Although, and as we discussed in section §3.1, 10μs of simulation time is nowhere near the unfolding times of Rop, the partial unfolding event that we did observe is not minor: the snapshot shown in Fig.5 corresponds —for the topmost turn alone— to the dissolution of three hydrophobic layers involving a total of 12 residues (these are the hydrophobic layers $Ala_{12}$-$Leu_{22}$-$Cys_{38}$-$Leu_{48}$ / $Cys_{52}$-$Gln_{34}$-$Leu_{26}$-$Ala_8$ / $Glu_5$-$Leu_{29}$-$Ala_{31}$-$Phe_{56}$). If a similar event were to take place at the other turn simultaneously, then only the central two hydrophobic layers (out of a total of eight) would have remained intact.

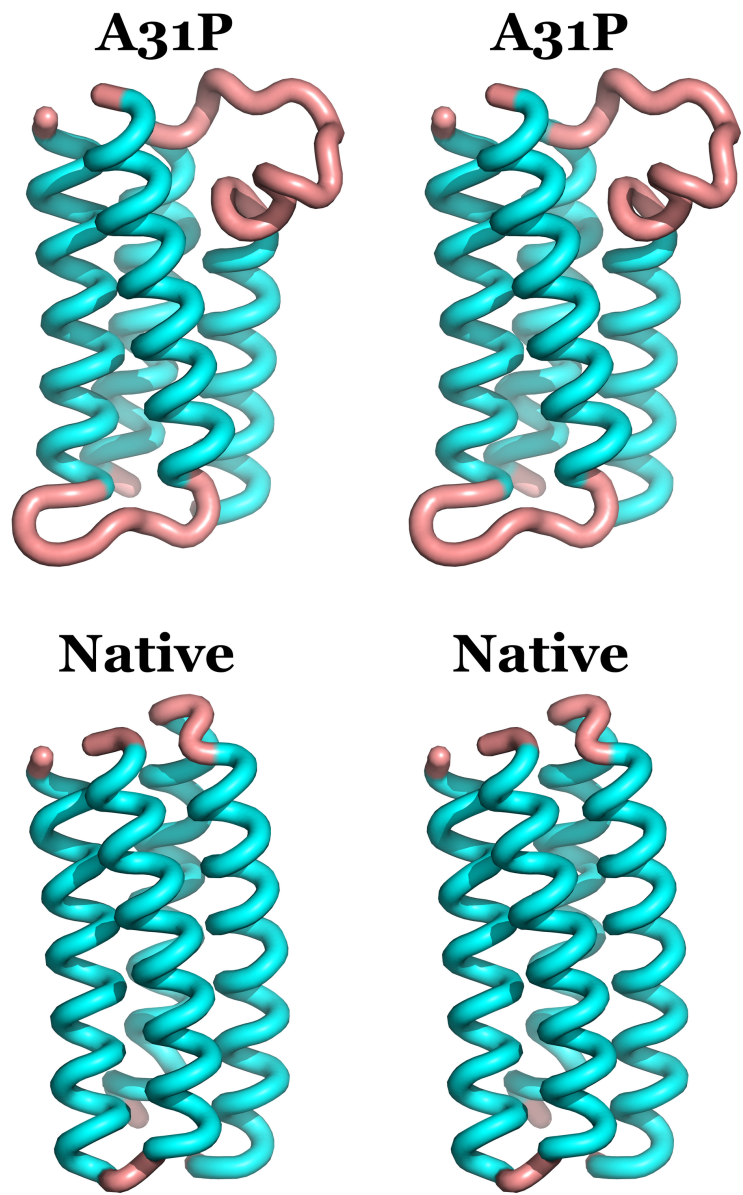

**Fig.5** Wall-eyed stereodiagrams of the Rop and A31P trajectory structures that differ the most from their respective starting conformations (see Fig.6 for the actual *RMSD vs. time* graphs). In this diagram the color coding indicates the STRIDE-derived[104] secondary structure assignment with cyan for α-helices, and salmon for coil/turns. The two structures have been superimposed on the helix that is closest to the viewer.

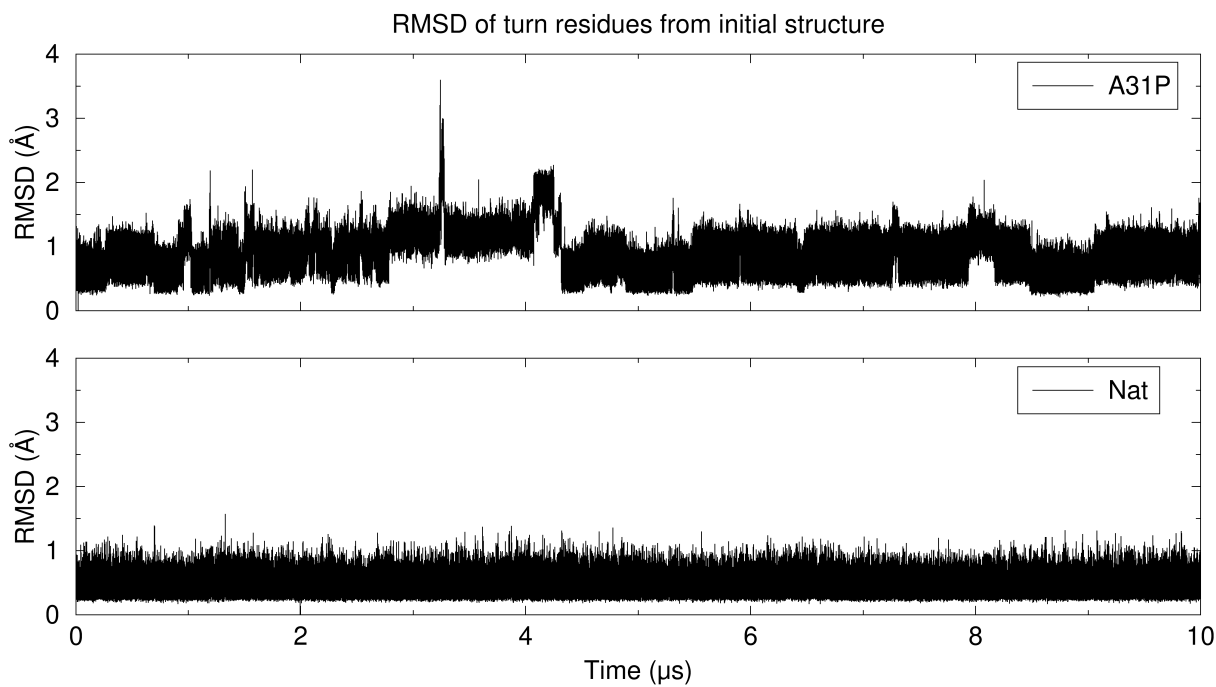

**Fig.6** RMSD from the initial structures for A31P (upper diagram) and native Rop (lower diagram) as a function of simulation time (in µs). The RMSDs were calculated using the Cα atoms of the turn residues only (24-39 inclusive). The structures corresponding to the maxima of these graphs are shown in Fig.5.

In sharp contrast with A31P, native Rop lived up to the expectations we had from a protein with a $T_m$ of more than 60°C : the great majority of the RMSD values shown in Fig.6 are less than 0.7Å, with an average of 0.52Å and a standard deviation of 0.097Å. Not unexpectedly, the structure shown in Fig.5 (lower panel) is for all practical purposes identical with the experimental Rop structure. We also note that all excursions to slightly higher RMSD values in Fig.6 are short lived and do not lead to a cooperative unfolding of the turns. Having said that, the RMSD from the starting structure is a very weak indicator of protein stability and carries no information about the presence (or otherwise) of correlated motion in the system under examination. For this reason we also present in the next section a comparison between the results obtained from a dihedral principal component (dPCA) analysis of the turns of the two structures.

## 3.5 Dihedral PCA detects correlated turn motion in A31P, but not in native Rop

We have analyzed the two trajectories using dihedral PCA[105–107] as implemented in the programs `carma`[87] and `grcarma`.[88] To be able to identify correlated motions of individual turns —a finding which could possibly indicate a cooperative unfolding of the A31P turn— we have performed this analysis using only the structures recorded from one of the two turns (the topmost one in Fig.5).

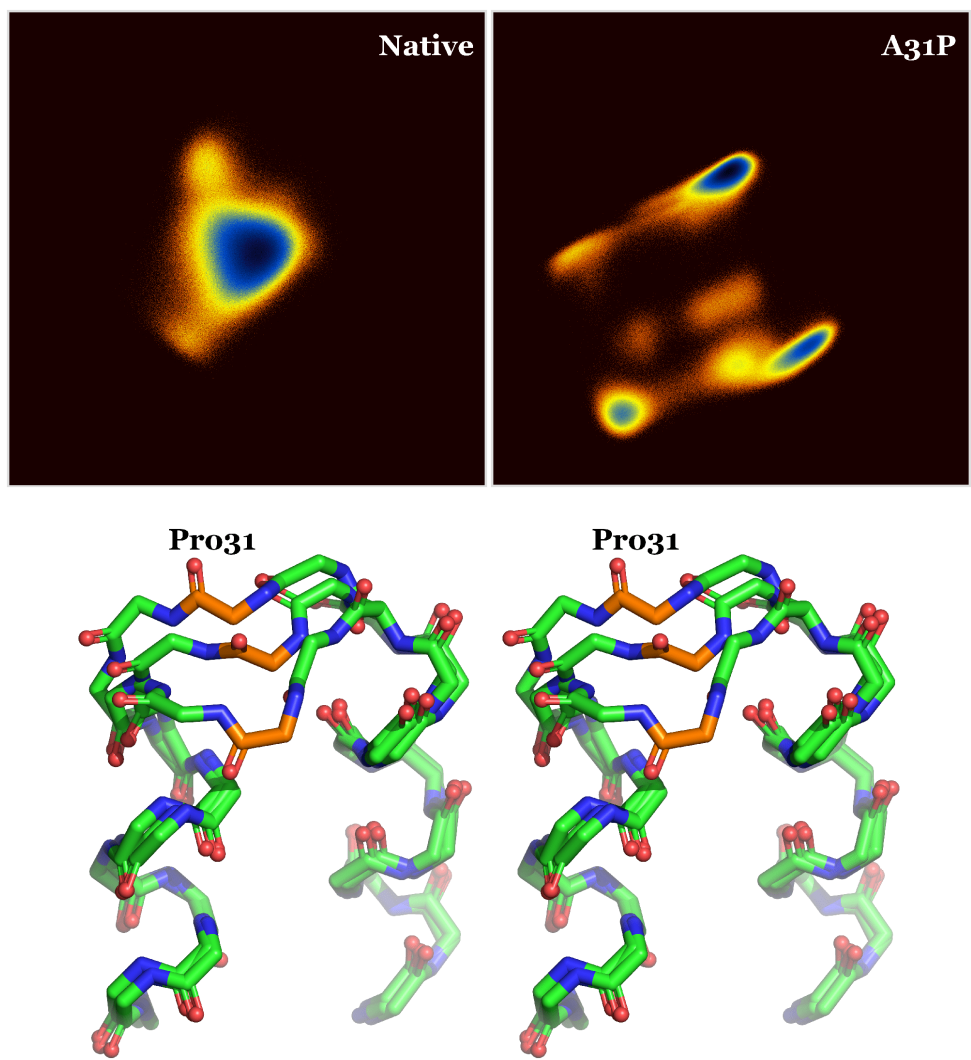

**Fig.7** Dihedral PCA and derived structures. The upper panel shows the log density distributions of the trajectory structures along the top two principal components for the native (left) and A31P (right) simulations. The distributions are on the same scale, and the color coding ranges from dark red (low density), through yellow, to dark blue (high density). The wall-eyed stereodiagram in the lower panel shows a superposition of three of the A31P conformers derived from the dPCA analysis. Only backbone atoms of a single turn region are shown, and $Pro_{31}$ is color marked and labeled.

The upper panel in Fig.7 shows the PC1-PC2 density distributions of the two trajectories (the two principal components used for this figure account for ~55% of the observed variance of the A31P trajectory, with the third component's contribution being at 4.3%). Starting from the native Rop structure, there is very little to say other than it is essentially harmonic with a single major conformational state that is practically identical with the crystallographic structure. A31P on the other hand demonstrates at least 7 distinct states, three of which are shown in the lower panel of Fig.7. Note that these states were derived from the whole trajectory, and that they do *not* correspond to the isolated unfolding event seen at $t \approx 3.2\mu s$ in Fig.6 (with its corresponding structure shown in Fig.5). As the structure diagram in Fig.7 illustrates, the turn of A31P undergoes significant and correlated fluctuations away from its starting structure, and away from the second monomer. An obvious model presents itself based on this figure: the A31P turn constantly fluctuates away from the sister monomer, creating a solvent accessible cavity, and exposing the hydrophobic core. Given that the top two hydrophobic core layers have contributions from atoms that belong to polar residues (like $Gln_{34}$ and $Glu_5$), it is not hard to imagine that intrusion of water molecules may stochastically destabilize the inner core and initiate unfolding, with the consequences shown in Fig.5.

# 4 Discussion

The aim of this communication was to answer the following question: why A31P does not fold like native Rop does ? The reason which makes this question meaningful and interesting lies with the results obtained from energy minimization, modeling, and structure prediction. Unanimously these methods indicated that a native-like A31P structure appears to be plausible and —computationally speaking— far more stable than the actual ('bisecting U') structure that A31P adopts. This paradox has, as far as we can tell, two solutions: either the mutation blocks the folding pathways leading to a native-like structure (thus making such a structure inaccessible during folding), or a native-like A31P structure is unstable and unfolds quickly. By performing two relatively long molecular dynamics simulations we obtained evidence in favor of the second solution: A native-like structure for A31P appears to be unstable and prone to unfolding as Fig.5 indicated. Dihedral PCA analysis of the trajectories offered addi-tional evidence supporting the model of an unstable A31P turn which fluctuates away from its sister monomer, exposes the protein's hydrophobic core, and nucleates unfolding events. These findings appear to make sense from a teleological point of view as well: after all, we already knew from crystallography that A31P does not fold like native Rop does. But such a pragmatic approach teaches us very little about the whys and wherefores of A31P and Rop folding.

To take these ideas a step further, we propose that the A31P mutation, and through the mechanism presented in section §3.5, destabilizes the turns of not just the native (*anti*) topology of Rop, but also of the *syn* topology previously observed in the $A_2I_2$ and $A_2L_2$ variants.[44,45,62,67] (structures 1F4M and 1F4N). If this proposition is accepted as plausible, then both of the two major and stable topologies of Rop and its variants are incompatible with the A31P mutation, which offers an explanation as to why (a) A31P is the only Rop variant with a completely new topology, and, (b) why the mutant's 'bisecting U' topology is only marginally stable. In a sense, Wolynes and coworkers[44] were prophetic in refusing to place A31P in the double-funneled energy landscape they showed in Fig.1 of their manuscript, mainly because this double-funneled energy landscape is probably non-existent in the case of A31P.

To summarize these ideas, our current model is the following: If for a Rop mutant or variant the *anti* or *syn* topology is accessible and structurally stable, then such a topology will be adopted and experimentally observed (with the provision that for some variants, both the *anti* or *syn* topologies may be populated). If, on the other hand, the mutation is incompatible with both of these two major topologies, then the protein explores its energy landscape with two possible outcomes. The first is that no other sufficiently stable conformation is accessible, in which case the protein will remain unfolded, and thus difficult to characterize structurally. The second possible (but rare) outcome, is that the specific mutation/variant under examination creates and populates a sufficiently stable third minimum in the energy landscape of Rop, stable enough to be observed experimentally. These mutation-induced high energy minima necessarily correspond to relatively unstable, molten-globule-like structures.[43] This presentation immediately suggests that the effect of the A31P mutation could in principle be reverted ('rescued') by a second mutation that would make the *anti* or *syn* topologies sta-ble and accessible again. Such attempts have been described by the Kokkinidis group[68] and include the D30P-A31G (PG) mutant, the D30G-A31P (GP) mutant and the D30P-A31P (PP) mutant. Of those, only the D30P-A31G mutant had a stable enough structure to be crystallized, and as it turned-out, its structure was native-like (which is not too surprising given that D30P is known to be native-like[54]).

The somewhat discouraging take-home message of all those experiments and calculations with mutants and variants of Rop, is that it is still very difficult to computationally predict what a mutation in the turn region of the protein will do. However, we do believe that sufficiently long molecular dynamics simulations could possibly help answer the following —limited, but useful— sub-question : "Will this turn mutant fold like native Rop, or not ?". If a native-like structure for the mutant is stable in the simulation, then this could be interpreted to mean that the *anti* topology is accessible and stable, and that the mutant will be native-like. Such simulations in the folded state could also help in the design of 'rescue' mutations that would revert the structural effects of A31P-like mutants. Parenthetically, such calculations can also be seen as an acid test for the accuracy of the current generation of force fields: can the simulations clearly show, for example, that the D30P turn mutant is stable and native-like, whereas A31P is unstable and incompatible with the native structure ? This is a question that we actively pursue presently.

This communication also highlighted the cautiousness with which modeling and energy minimization exercises must be treated. All indications from the examination of a native-like structure of A31P suggested that such a structure appeared to be entirely plausible and stable. And all indications were wrong. We believe that at least part of the problem lies with what could be called 'enthalpy bias'. We suggest that both human observers and most energy minimizers underestimate the contributions from the more difficult to visualize and calculate entropy contributions, involving both the protein and the surrounding solvent. We strongly suspect that both humans and the energy minimizers correctly identified that a native-like A31P structure is indeed the most stable conformation based on the enthalpy contributions, but they completely missed the entropic terms' contributions, which by all appearances are responsible for the A31P unfolding.

We should like to close with this discussion with a summary of what we perceive to be the most important limitations of this work. The first is that the presentation in the previous paragraphs may leave the impression that we perceive this work as establishing beyond reasonable doubt that a native-like A31P structure is unstable. This is definitely not the case. To start with, we have not observed a complete unfolding event, only a short-lived partial unfolding of the turn region. Additionally, the limited evidence that we did obtain is purely computational, with no direct experimental verification (which, admittedly, would be exceedingly difficult to obtain), and wholly based on a given force field with its ever present approximations and limitations. The second serious limitation we perceive concerns the time scale of this analysis. Compared with the folding times observed in the Rop family, 10 μs is so little, that we can not even establish beyond doubt that even native Rop is as stable as its simulation indicated. The third limitation is that due to the complexity of system, most of our treatment is qualitative and in some cases completely descriptive. For example, why is it that only A31P (but not, for example, the D30G-A31P mutant) stabilizes the 'bisecting U' topology ? The answer to such questions would have required the faithful mapping of the corresponding folding landscapes, and this is not feasible with present day computing capabilities. And, thus, after more that 40 years of intense studying, we must still conclude that the Repressor of Primer protein successfully eludes our attempts to quantitatively understand its folding and how it determines its structure.